\documentclass[12pt]{iopart}

\pdfoutput=1

\usepackage{iopams}
\usepackage{graphicx} 
\usepackage{color}
\usepackage{soul}
\usepackage{url}
\usepackage{cite}
\usepackage[colorlinks]{hyperref}

\usepackage[caption=false]{subfig}

\hypersetup{pdftitle=Emergence of giant strongly connected components in continuum disk-spin percolation, pdfauthor={Francesco Caravelli, Marco Bardoscia, Fabio Caccioli}}

\begin{document}

\title[Emergence of giant SCC in continuum disk-spin percolation]{Emergence of giant strongly connected components in continuum disk-spin percolation}

\author{Francesco Caravelli$^{1,2}$, Marco Bardoscia$^2$, Fabio Caccioli$^{3,4}$}

\address{$^1$ Invenia Labs, 27 Parkside Place, CB1 1HQ Cambridge, UK}
\address{$^2$ London Institute for Mathematical Sciences, 35a South St.\ W1K 2XF London, UK}
\address{$^3$ Department of Computer Science, University College London, Gower St.\ WC1E 6BT London, UK}
\address{$^4$ Systemic Risk Centre, London School of Economics and Political Sciences, London, UK}
\eads{\mailto{francesco.caravelli@invenialabs.co.uk}, \mailto{marco.bardoscia@gmail.com}, \mailto{f.caccioli@ucl.ac.uk}}

\begin{abstract}
We propose a continuum model of percolation in two dimensions for overlapping disks with spin. In this model the existence of bonds is determined by the distance between the centers of the disks, and by the scalar product of the (randomly) directed spin with the direction of the vector connecting the centers of neighboring disks. The direction of a single spin is controlled by a ``temperature'', representing the amount of polarization of the spins in the direction of an external field. Our model is inspired by biological neuronal networks and aims to characterize their topological properties when axonal guidance plays a major role. We numerically study the phase diagram of the model observing the emergence of a giant strongly connected component, representing the portion of neurons that are causally connected. We provide strong evidence that the critical exponents depend on the temperature.
\end{abstract}

\maketitle

\section{Introduction} \label{sec:intro}
Percolation theory is one of the simplest models exhibiting a phase transition \cite{orig,percol}. Due to its simplicity, it provides fundamental insights that facilitate the understanding of many phenomena across biology, physics, and geophysics \cite{percol,percol2,percol3}, and is recently experiencing a revival of interest \cite{exp1,exp2,Noh2004,contperc1,contperc2,contperc3,rep}. In this paper we use the theory of percolation to study the properties of a random spatial network model inspired by axonal guidance in neuronal networks. 

The central nervous system is an intricate, disordered network composed of tiny processing units, the neurons, connected to each other in a complex manner \cite{cell}. Neurons themselves have a rather complicated dynamics \cite{neurdyn} and are composed mainly of three parts: \textit{dendrites}, which provide the input signals, a central body called \textit{soma}, and the \textit{axon}, which transmits the output and can reach up to one meter in length \cite{cell}. Most models of neuronal dynamics \cite{neurdyn,intfire,HH,FitzHugh, Nagumo} agree that the topology of the connections between neurons plays an important role. As a consequence, several possible network structures have been investigated, ranging from fully connected graphs \cite{Jaeger, Maass} 
with homogeneous or random couplings \cite{Kandel} to graphs in which the connectivity depends on the distance between neurons \cite{RandomDistance}. 

\begin{figure}
\centering
\begin{tabular}{c}
\subfloat[\ ]{
\includegraphics[scale=0.4]{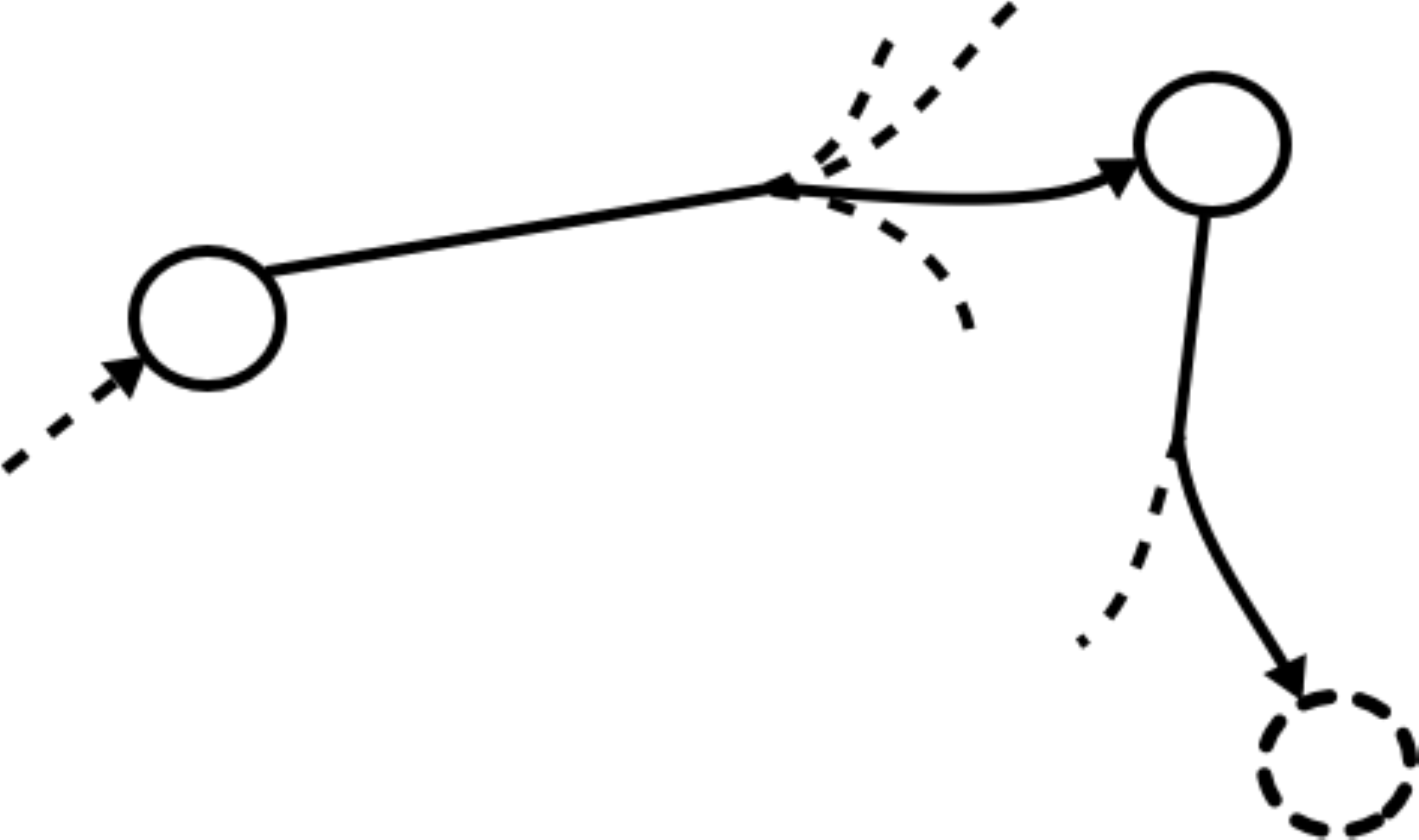}
\label{fig:example1}
} \\
\subfloat[\ ]{
\includegraphics[scale=0.4]{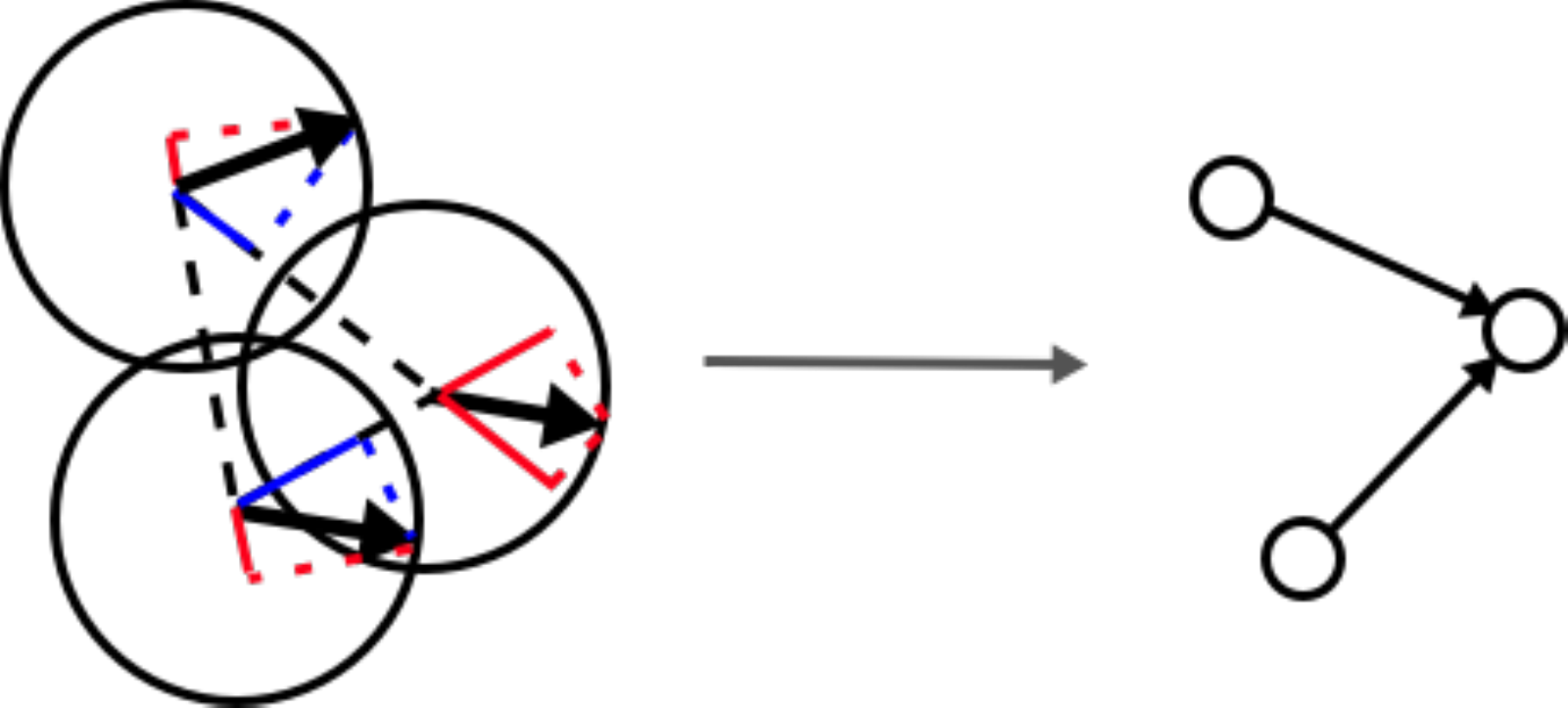}
\label{fig:example2}
}
\end{tabular}
\caption{A graphical representation of the model we consider in this paper. \textit{(a)} Schematized structure of real neuronal networks. The circle represents the soma, meanwhile the lines represent axons branching out from neuronal cells along particular directions. \textit{(b)} Equivalent description in terms of overlapping disks of radius $\tilde{p}$ equipped with a  spin. The circle represents the maximum extension of the axon, the spin represents its direction and the dashed lines connect the centers of the neurons where the soma is located. The blue lines represent a positive projection, while the red lines a negative projection. On the right, we show the resulting directed graph.}
\label{fig:example}
\end{figure}

The development of the mammal brain is clearly very complex, starting from the neural plate and following several steps. It is characterized by an initial exponential growth driven by neuronal migration, differentiation, and axonal guidance \cite{growth}. After its formation, in the years of maturity, the number of neural connections in all areas of the brain dramatically decreases, a process known as pruning \cite{pruning}.
In this paper we abstract from such complexity and study a simple, stylized model to describe how the formation of a giant cluster of neurons depends on a few parameters that in principle could be measured experimentally. The mechanism we propose for the formation of a bond between two neurons takes two effects into account. First, in the neocortex the majority of connections between neurons are short-range (with only few connections that make the network small-world \cite{Lefort}), so it is reasonable, at least as a first approximation, to connect only neurons that are close to each other. Second, since the axon is much longer than the size of the body of the neuron, it is clear that the direction of the axon must play an important role \cite{fg}. Therefore, we allow a neuron to be connected only with other neurons that fall within a certain angle around the direction of its axon. Our simplified characterization of a neuron is shown in figure \ref{fig:example}. Neurons are represented as disks equipped with a spin: the radius of the disks accounts for the finite size of neurons, while the spin for the directionality of the axon. 

On one hand, the proposed model falls in the category of a continuum (two-dimensional) disk percolation model, which has been studied in a variety of contexts and also for different shapes as an alternative to disks \cite{Stanley,gsdt,Quintanilla,powerlaw}. However, an important difference is that we account for the presence of directionality in the formation of bonds (due to the direction of the axon), meaning that the underlying network is directed. On the other hand, models of directed percolation have been previously investigated both on a lattice to explain the flow in porous materials \cite{dp1,cardy,dp2,dp3} and on complex networks \cite{pren2,pren1,prln1,net1,net2}. In our case disks are placed in a euclidean space, which is in stark contrast both with the former, characterized by the regular geometry of a lattice, and with the latter, in which there is no embedding in a metric space.  
Therefore, our model can be put at the intersection between continuum percolation, directed percolation, and random networks, and studying its behavior is interesting in itself.

The presence of a giant percolating cluster is particularly relevant in view of the aforementioned pruning process, which can be depicted as a removal of bonds between neurons. Since the underlying network is directed, one could in principle choose different definitions for the giant, percolating cluster of disks. We focus on the emergence of a \emph{basin of influence} in which every neuron belonging to the giant cluster can propagate a signal, i.e.\ information within the basin. Therefore, it represents the set of neurons which are mutually causally connected and, incidentally, it mirrors the definition of \emph{fully recurrent} neural networks. 

Such property is also important in the context of artificial neural networks \cite{Hopfield}. The recent success of deep learning algorithms \cite{Hinton} suggests that a layered feedforward (i.e.\ acyclic) structure inspired by the topological structure of realistic neuronal cell networks in the visual cortex \cite{hartley} can significantly improve the performance of artificial neural networks. Thus, one might argue that the presence of a hierarchical structure is key for performing fast learning artificial tasks. From this perspective, the existence of giant strongly connected cluster is detrimental, as it rules out the presence of a layered feedforward structure.

The paper is organized as follows. We first introduce the model and the relevant definitions in section \ref{sec:model}. In section \ref{sec:analysis} we analyze the model numerically and explore their universality classes by means of finite-size scaling. Conclusions follow in section \ref{sec:conclu}.

\section{The model} \label{sec:model}
We consider a simple, stylized model to describe the mechanism governing the formation of a giant cluster of neurons. We represent a neuron as a disk with a given radius. The region inside the disk encloses the soma, the dendrites, and the axon. To each disk we also associate a direction (a unit vector which we call \emph{spin}) that models the spatial orientation of the axon. We then choose a simple set of rules to decide if two disks belong to the same cluster. We start with implementing a proximity constraint, meaning that two disk can be part of the same cluster only if they overlap\footnote{Long-range connections can be accounted for by introducing a random rewiring, in the spirit of small-world network models \cite{wattsstrogatz}.}. However, even if two neurons are close, they might not connect, because of the orientations of their axons. Therefore we add the following directional constraint: we form the link $i \rightarrow j$ between two overlapping disks $i$ and $j$ if the center of disk $j$ is in the half-plane spanned by an angle $\phi$ around the direction of disk $i$. Intuitively, it means that looking in the direction of disk $i$ we can detect the center of disk $j$ within an angle $\phi$, i.e.\ in a limited range of angles in front of us. We can see that such  prescription is not symmetric: the existence of the link $i \rightarrow j$ does imply the existence of the reverse link $j \rightarrow i$. The directionality in the process that leads to the creation of the network is intended to mirror the physiological directionality of signals in neuronal networks, which are transmitted from the soma through the axon.
Such problem naturally falls within the framework of random geometric graphs. More formally, we consider a two-dimensional square box whose sides have length $L = 1$ in which we drop $N$ disks at random locations. We denote with $\vec{x}_i$ the spatial coordinate of disk $i$ and with the unit vector $\hat{s}_i$ its spin. All disks have radius $\tilde{p}$. We now build a directed graph associating to each disk a node in the graph and using the aforementioned rules to draw edges between nodes. Denoting with $A_{ij}$ the element $(i, j)$ of the adjacency matrix of the graph, we can write both the proximity and the directional constraint in the following compact form:
\begin{equation}
A_{ij} = \Theta(2\tilde{p} - |\vec{x}_j - \vec{x}_i|) \; \Theta \left( \hat{s}_i \cdot \frac{\vec{x}_j - \vec{x}_i}{|\vec{x}_j - \vec{x}_i|} - \cos\frac{\phi}{2} \right) \, ,
\end{equation}
where $\Theta(x)$ is the Heaviside function, i.e.\ $\Theta(x) > 0$ for $x > 0$, and $\Theta(x) = 0$ otherwise, and $|\cdot|$ is the vector square root of 2-norm. We can see that the first $\Theta$ implements the proximity constraint: It is different from zero only if the vector $\vec{x}_j - \vec{x}_i$ (which connects the center of disk $i$ with the center of disk $j$) is shorter than $2\tilde{p}$, the diameter of a disk, i.e.\ only if the two disks overlap. The second $\Theta$ implements the directional constraint. To see it, it is sufficient to note that the scalar product $\hat{s}_i \cdot \frac{\vec{x}_j - \vec{x}_i}{|\vec{x}_j - \vec{x}_i|}$ is simply equal to the cosine of the angle between the spin $\hat{s}_i$ and the unit vector connecting the center of disk $i$ with the center of disk $j$. Therefore, this term is larger than zero precisely when such angle is in the interval $[-\phi/2, +\phi/2]$.

Let us now mention a crucial result in the context of random graphs: Let us scatter $N$ balls of radius $p/{N}^{1/d}$ in an hypercube of $d$ dimensions and linear size 1, and let us link two balls if they overlap. It has been proven \cite{Penrose2003} that, for $d \geq 2$ and in the limit $N \rightarrow \infty$, there exists $p_c \in (0, \infty)$ such that for $p > p_c$ almost surely an infinite cluster of balls appears, while for $p < p_c$ almost surely it does not, a behavior that closely resembles a percolation transition. Since our model generalizes the aforementioned mechanism by taking into account the spin direction, it is reasonable to expect a similar transition to occur. Therefore, we adopt the same scaling $\tilde{p} = p/\sqrt{N}$ for the radius of disks and we define $p_c$ as the normalized radius of disks at which an infinite cluster of disks appears in the limit $N \rightarrow \infty$. We also introduce $S(p)$, the average cluster size for a given value of $p$ (excluding the largest ``infinite'' cluster \cite{Rintoul1997}), $P(p)$, the probability that an arbitrary disk belongs to the infinite cluster for a given value of $p$, and $\xi$ the correlation length\footnote{$\xi$ is defined from $g(\mathbf{r}) = e^{-r/\xi}$, where $g(\mathbf{r})$ is the correlation function, i.e.\ the probability that a point identified by the vector $\mathbf{r}$ applied to a point in a finite cluster belongs to the same finite cluster.}. Close to the critical point $p_c$, such quantities are expected to behave as power laws in the variable $t = p - p_c$:
\numparts
\begin{eqnarray}
&S(p) \simeq |t|^{-\gamma} \qquad &p \rightarrow p_c \\
&P(p) \simeq t^{\beta} \qquad &p \rightarrow p_c^+ \\
&\xi \simeq |t|^{-\nu} \qquad &p \rightarrow p_c \, .
\end{eqnarray}
\endnumparts

Finally, let us point out that rescaling the radius ($\tilde{p} = p/\sqrt{N}$) and keeping the linear size of the box fixed ($L=1$) is equivalent to keeping the radius constant ($\tilde{p} = p$) and rescaling the box ($L = \sqrt{N}$), which is  more intuitive when dealing with quantities such as the correlation length. 

It is worth to remark that in our case edges have a direction. On a directed graph there are several possible definitions of clusters, corresponding to different kinds of connectedness. Following \cite{newman}, we identify clusters with \emph{strongly connected subgraphs}.  More precisely, given a \emph{directed} graph $\mathcal{G}=\{V,E\}$, being $V$ the set of nodes of cardinality $N$ and being $E$ the set of directed edges with maximal cardinality $N(N-1)$, a strongly connected subgraph $\mathcal{G}^\prime$ of $\mathcal{G}$ is a collection of nodes and directed edges of $\mathcal{G}$, such that there exists a directed path between each pair of vertices in $\mathcal{G}^\prime$ (i.e.\ in both directions). A \emph{maximally} strongly connected subgraph is the maximal subgraph that is strongly connected. Maximally strongly connected subgraphs will serve as our definition of percolating clusters. 

\begin{figure}
\centering
\includegraphics[width=0.49\columnwidth]{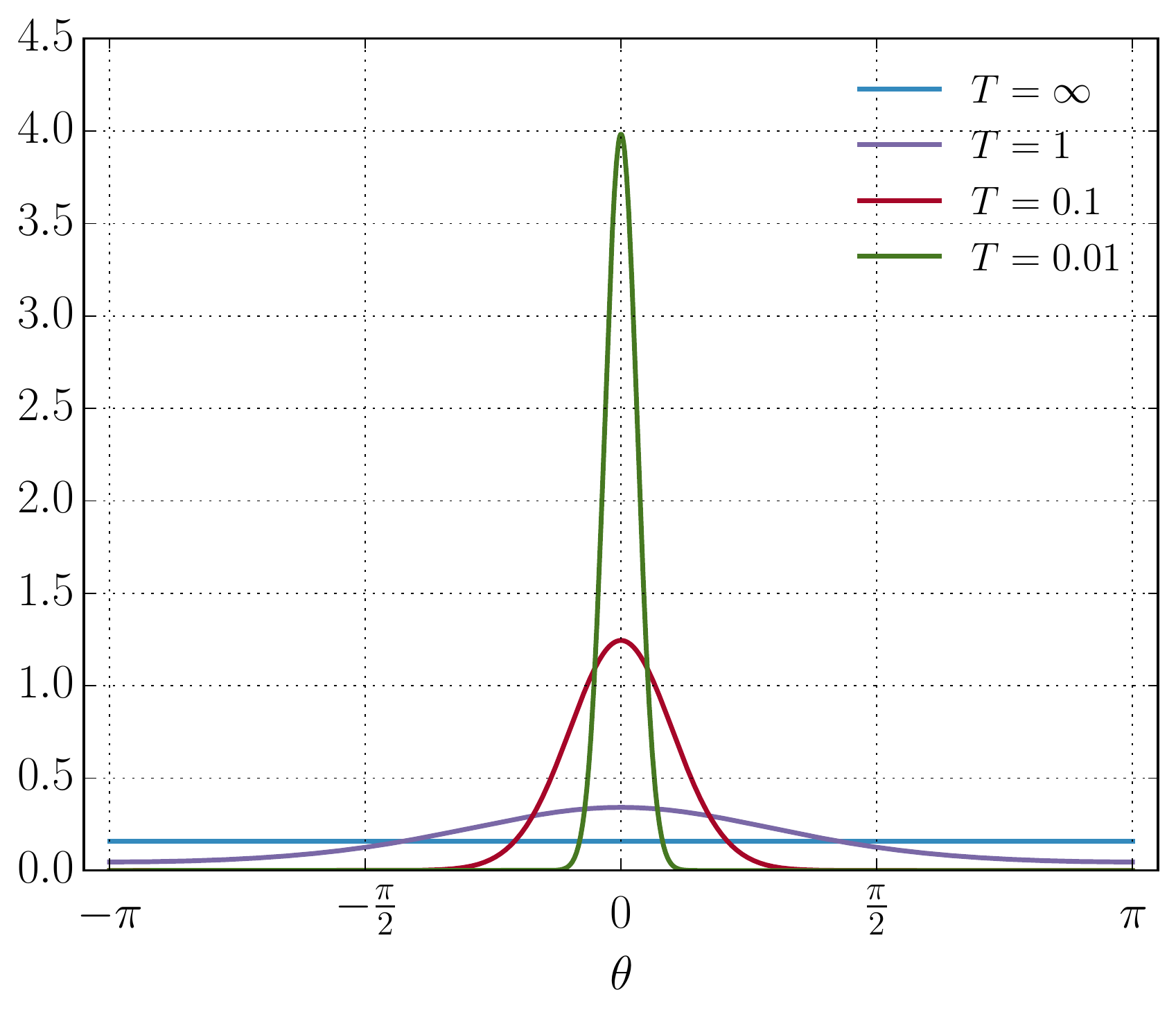}
\caption{ Probability distributions of $\theta$, the angle that spins form with the x axes for different values of $T$, the temperature controlling the fluctuations around such direction.}
\label{fig:spin_dis}
\end{figure}

\begin{figure}
\centering
\includegraphics[width=0.49\columnwidth]{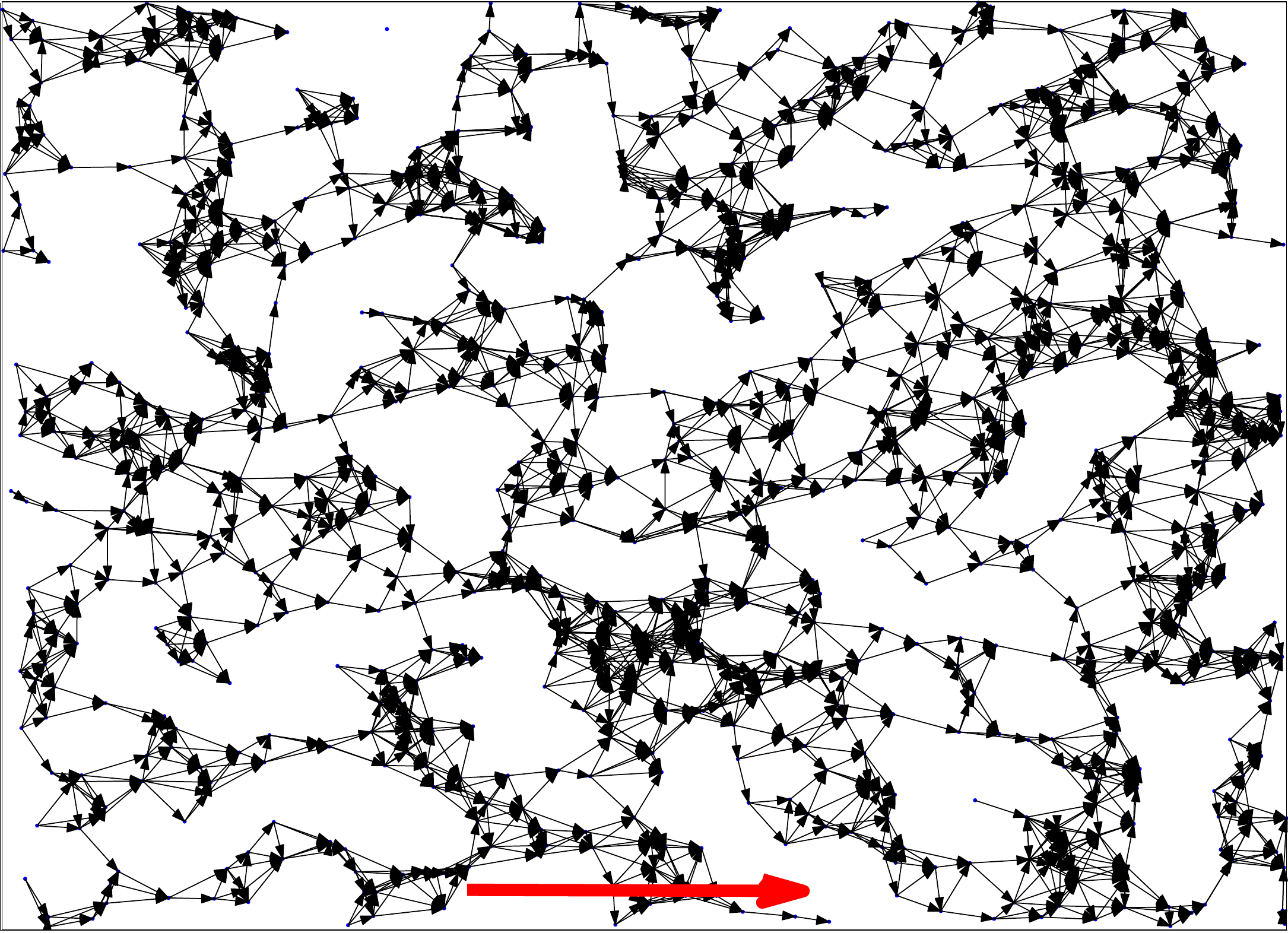}
\caption{Single realization of random positions and orientations of 100 disks with an external field pointing in the direction of the red arrow for $1/T=0.1$.}
\label{fig:disex}
\end{figure}

So far we have not specified any distribution for the direction of spins, which is defined by the angle $\theta_i$ that $\hat{s}_i$ forms with the horizontal side of the box. We choose a general approach in which a single parameter allows us to interpolate between the case in which all spins are aligned along a privileged direction and the case in which the angles $\theta_i$ are uniformly distributed in  in the interval $[-\pi, \pi]$. The rationale of such choice is to provide a unified framework encompassing both feed-forward networks (resembling the visual cortex \cite{lamme}), in which connections have a strong directionality, and recurrent networks \cite{sporns,janoos}, in which cycles appear frequently.
Without loss of generality, we choose a privileged direction defined by the unit vector $\hat{h}$ parallel to the horizontal side of the box and we fix such direction \emph{on average}, without assuming any further knowledge on the probability distribution of $P(\theta_i)$. Hence $P(\theta_i)$ will be the maximum entropy distribution compatible with $\hat{s}_i \cdot \hat{h}$ being fixed on average:
\begin{equation}
P(\theta_i) = \frac{1}{Z} e^{\frac{1}{T} \hat{s}_i \cdot \hat{h}} = \frac{1}{Z} e^{\frac{1}{T}\cos \theta_i} \, ,
\end{equation}
$T$ being the Lagrange multiplier controlling the fluctuations around the chosen direction and $Z$ being a normalization factor, which is readily determined:
\begin{equation} \label{eq:partfuncspdisk}
Z = \int_0^{2\pi} P(\theta_i) d \theta_i= \int_0^{2\pi} e^{\frac{1}{T}\cos \theta_i} d \theta_i= 2 \pi I_0 (1/T) \, ,
\end{equation} 
where $I_0(x)$ is the modified Bessel function of the first kind. In figure \ref{fig:spin_dis} we show $P(\theta_i)$ for different values of $T$. For small values of $T$ the distribution is concentrated around $\theta_i = 0$, the privileged direction. As $T$ increases, fluctuations around such direction increase ($\langle \theta_i ^ 2 \rangle$ is an increasing function of $T$), and for $T \rightarrow \infty$ the uniform distribution is recovered. A realization of random positions and orientations of 100 disks (disks boundaries are not shown to enhance readability) is shown in figure \ref{fig:disex}. 

\section{The numerical analysis} \label{sec:analysis}
We perform extensive numerical simulations\footnote{In order to generate random numbers we use the GNU Scientific Library (GSL) \cite{gsl}. The computation of the size of strongly connected components of the resulting graphs took advantage of the Boost Graph Library \cite{boost}, which implements Tarjan's algorithm \cite{tarjan}.} to characterize the percolation transition. We compute numerical estimators of the strength of the infinite cluster $P(p)$ and of the average cluster size $S(p)$. $P(p)$ is estimated simply as the fraction of disks in the largest cluster, while $S(p)$ is estimated as the average of all cluster sizes, after the removal of the largest one. We generate multiple realizations of the random locations of disks within the box,  and by averaging both quantities over such realizations we define $\Delta(p) = \langle P(p) \rangle$ and $\chi(p) = \langle S(p) \rangle$. We use open boundary conditions. Periodic boundary conditions would not simplify our numerical analysis, but could introduce a bias towards larger SCCs, because of the edges formed across the boundary. Such bias would be especially strong when the process of the formation of edges is highly directional. In fact, either for $\phi \to 0$ or for $T \to 0$ the cycles needed to form a SCC can be only formed through the boundary.

\begin{figure}
\centering
\includegraphics[width=0.49\columnwidth]{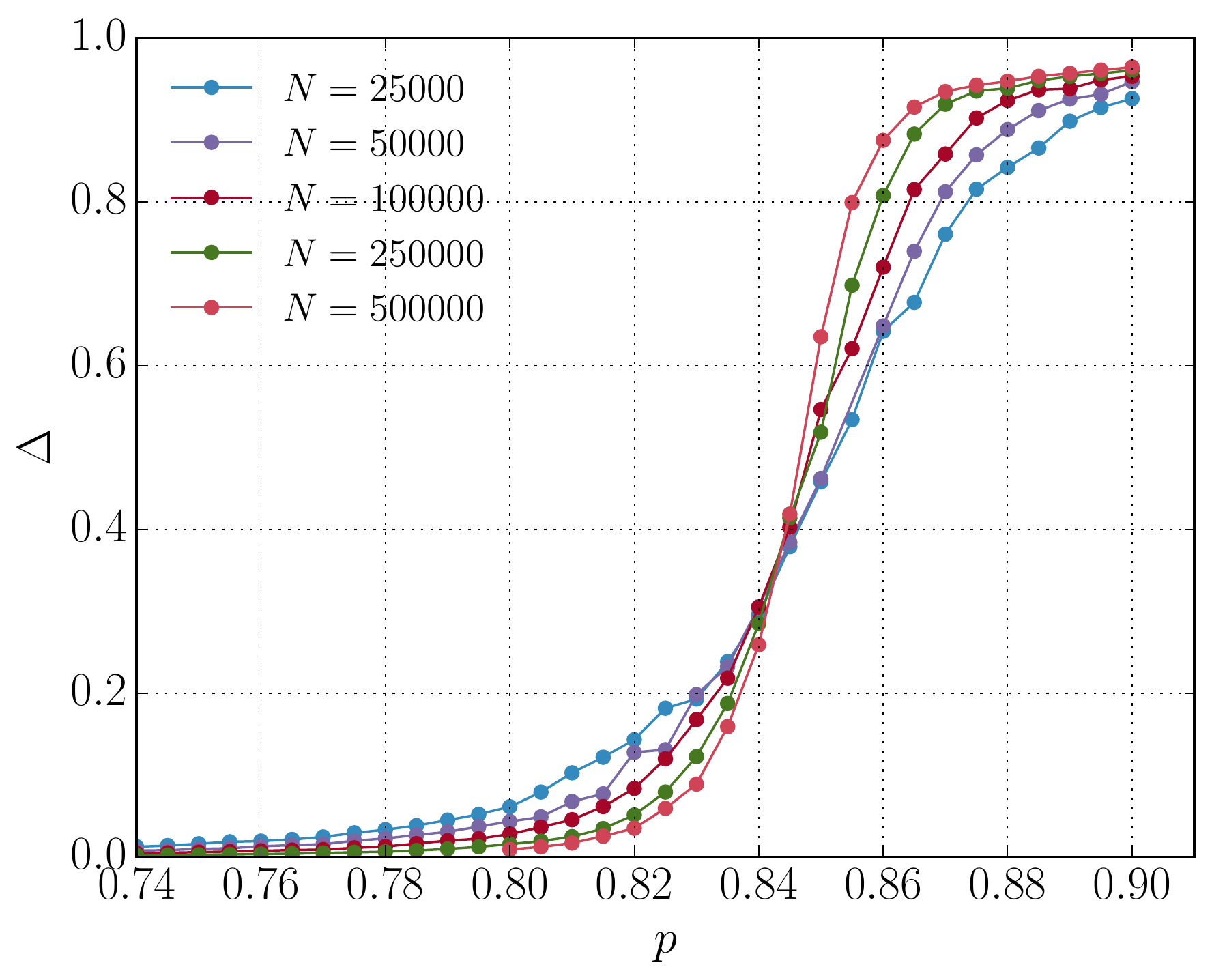}
\includegraphics[width=0.49\columnwidth]{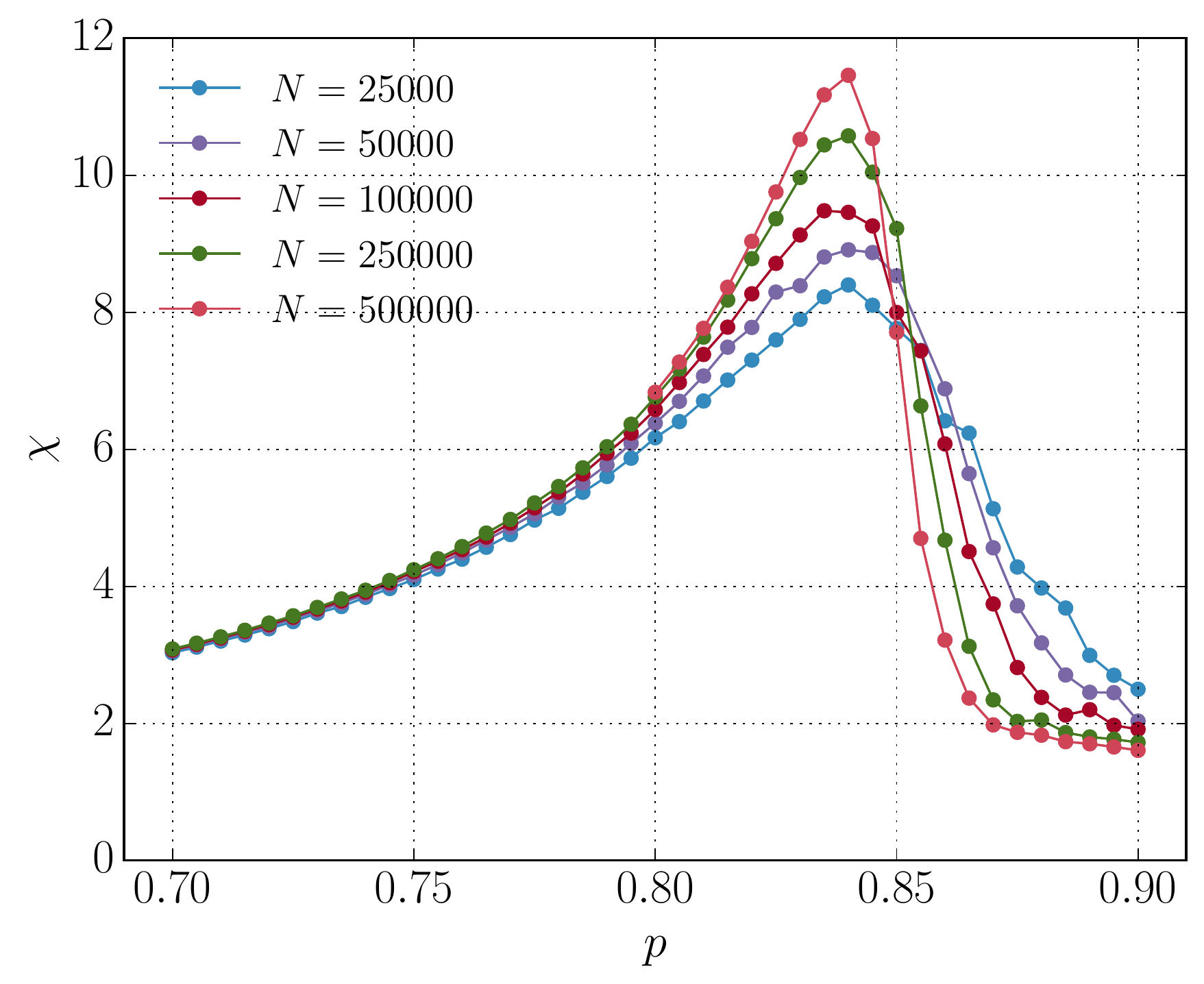}
\caption{$\Delta$, fraction of disks in the largest cluster (\textit{left}) and $\chi$, average cluster size (\textit{right}) as a function of $p$, the normalized radius of disks. Every point is averaged over 100 realizations, $T = 1$, and $\phi = \pi$. $\Delta$ goes from zero to one as $p$ increases and the transition becomes sharper and sharper as the number of disks $N$ increases. The peak $\chi^*$ of the average cluster size corresponds to the appearance of a GSCC (see main text).}
\label{fig:largest_and_average}
\end{figure}

Let us focus initially on single values of $T$ and $\phi$, which we set respectively equal to one and $\pi$. In figure \ref{fig:largest_and_average} we plot $\Delta(p)$ and $\chi(p)$ as functions of $p$, the normalized radius of disks. We see that for small values of $p$, $\Delta(p)$ is very close to zero, while for large values of $p$, $\Delta(p)$ approaches one, signaling the presence of a giant strongly connected component (GSCC). The transition between the regimes becomes sharper and shaper as the size of the system increases. As regards $\chi(p)$, for small values of $p$ the average cluster size is slightly above one, showing that all disks form isolated clusters. As $p$ increases, $\chi$ reaches the peak value $\chi^*$ and then decreases. The reason is that the average cluster size is computed by excluding the largest cluster, whose size diverges in the limit $N \rightarrow \infty$. In a system of finite size, the appearance of the GSCC has the effect of reducing the size of the other clusters, effectively leading to a decrease of $\chi$. As a consequence, the position $p^*$ of the peak $\chi^*$ is a good estimator of the critical value of $p$ for a system of finite size.

\begin{figure}
\centering
\includegraphics[width=0.49\columnwidth]{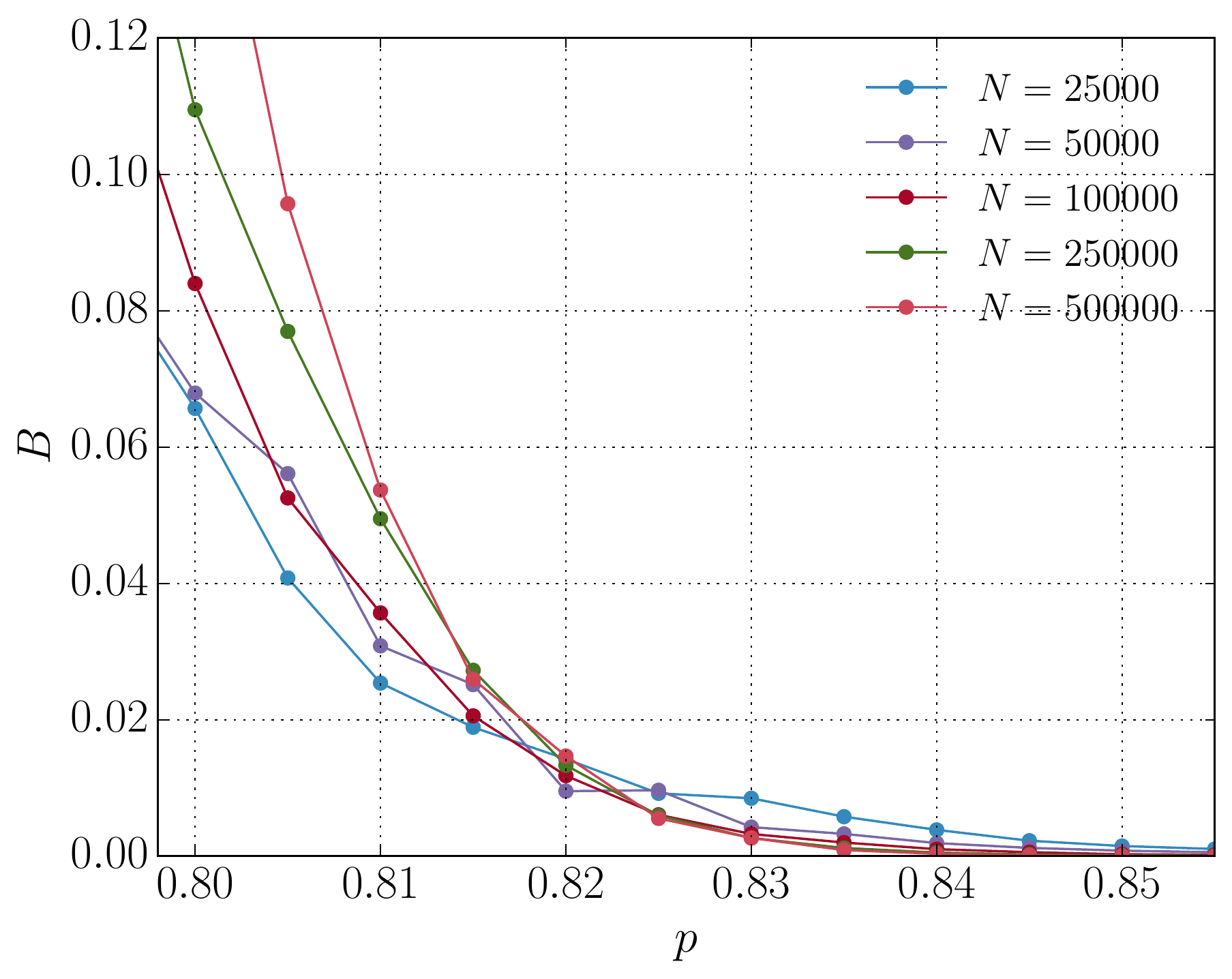}
\includegraphics[width=0.49\columnwidth]{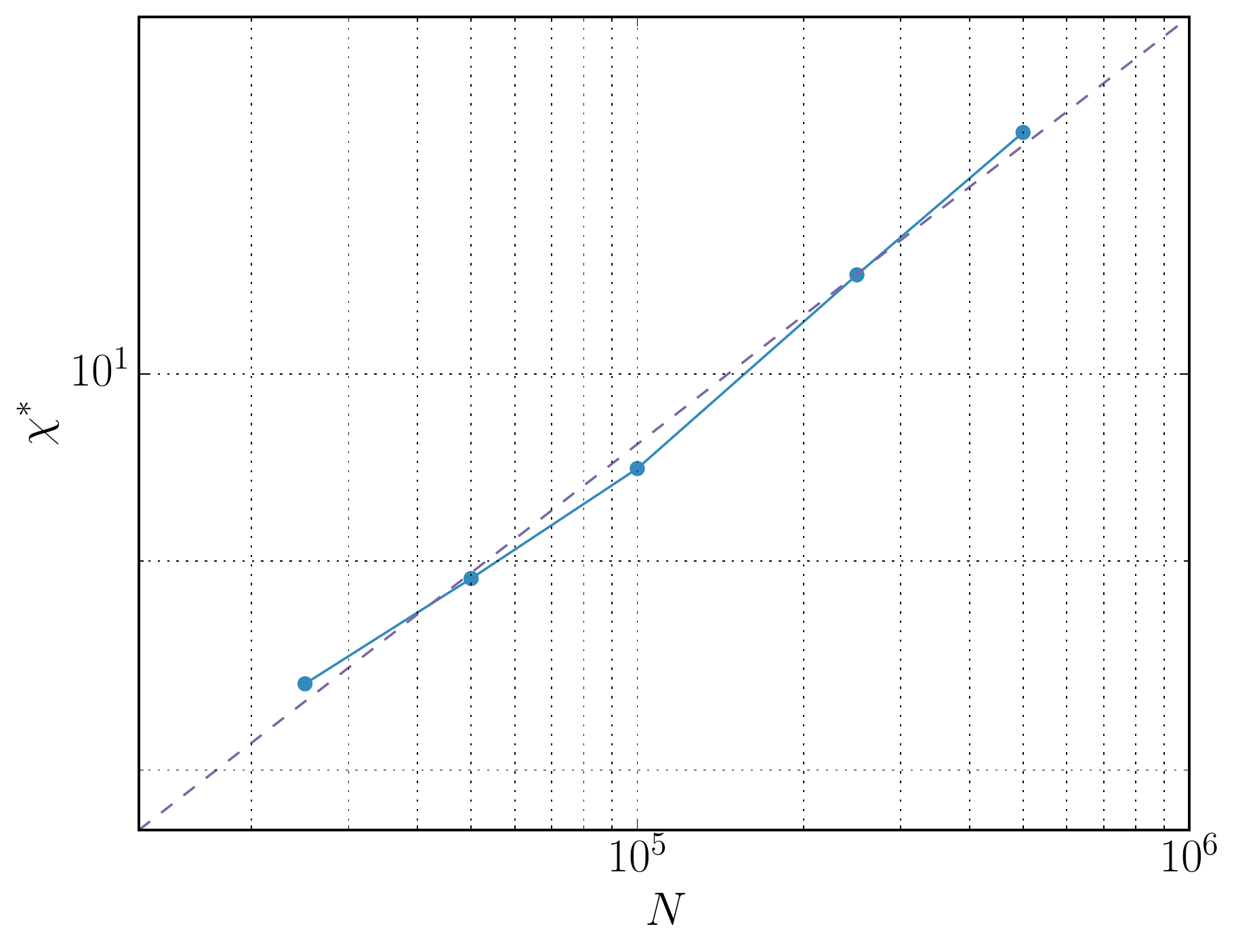}
\caption{\textit{Left:} $B$, Binder cumulant with respect to $p$, the normalized radius of disks, for different values of $N$, the number of disks. Every point is averaged over 100 realizations, $T = 1$, and $\phi = \pi$. All curves intersect approximately at $p = 0.82$, which is therefore the estimate for the critical point. \textit{Right:} finite-size scaling to estimate the ratio $\gamma / \nu$ for $T = 1$ and $\phi = \pi$. Log-log plot of $\chi^*$, the peak value of the average cluster size with respect to $N$. The linear fit is good, yielding $\gamma / \nu = 0.21 \pm 0.03$, with $R^2 = 0.9945$.}
\label{fig:binder_and_gammanu}
\end{figure}

In order to estimate the critical value of $p$ we use the Binder cumulant method \cite{binder}. The idea is to build a quantity that \emph{at the critical point} does not depend on the size of the system. The important observation is that in system of finite size at the critical point the correlation length $\xi$ is approximately the linear size of the system $\xi \simeq L$, from which it follows that $t \simeq L^{-1/\nu}$, $\Delta(p_c) \simeq L^{-\beta/\nu}$, and $\chi(p_c) \simeq L^{\gamma/\nu}$. If we now build the quantity
\begin{equation}
B(p) = \frac{\chi(p)}{\Delta(p)^2 L^d} \, ,
\end{equation}
we have that $B(p_c) \simeq L^{\gamma/\nu + 2\beta/\nu - d}$. Using the hyperscaling relation $2\beta + \gamma = d \nu$, we can see that $B(p_c)$ does not depend on $L$. Therefore, if we plot $B(p)$ vs $p$ for different sizes of the system all curves will cross at $p_c$. For illustrative purposes we focus again on $T = 1$ and $\phi = \pi$ and, from the left panel of figure \ref{fig:binder_and_gammanu} we see that the curves corresponding to different sizes of system cross each other approximately at $p = 0.82 \pm 0.005$.

In order to estimate the ratio $\gamma / \nu$ we use finite-size scaling. Let us recall that $\chi(p_c) \simeq L^{\gamma/\nu} \simeq N^{\gamma/2\nu}$, hence diverging as $L \rightarrow \infty$. For a system of finite size, the value $\chi^*$ corresponding to the peak in a plot of $\chi(p)$ vs $p$ can be used as an estimator of $\chi(p_c)$. Hence, a log-log plot of $\chi^*$ vs $N$ should yield points disposed on a straight line with slope equal to $\gamma / 2\nu$. The error on the ratio $\gamma / \nu$ can be computed simply by propagating the error on the linear fit. From the right panel of figure \ref{fig:binder_and_gammanu} we can see that for $T=1$ and $\phi=\pi$ the linear fit is satisfactory, resulting in the following estimate of the ratio: $\gamma / \nu = 0.21 \pm 0.03$, with $R^2 = 0.9945$.

\begin{figure}
\centering
\includegraphics[width=0.49\columnwidth]{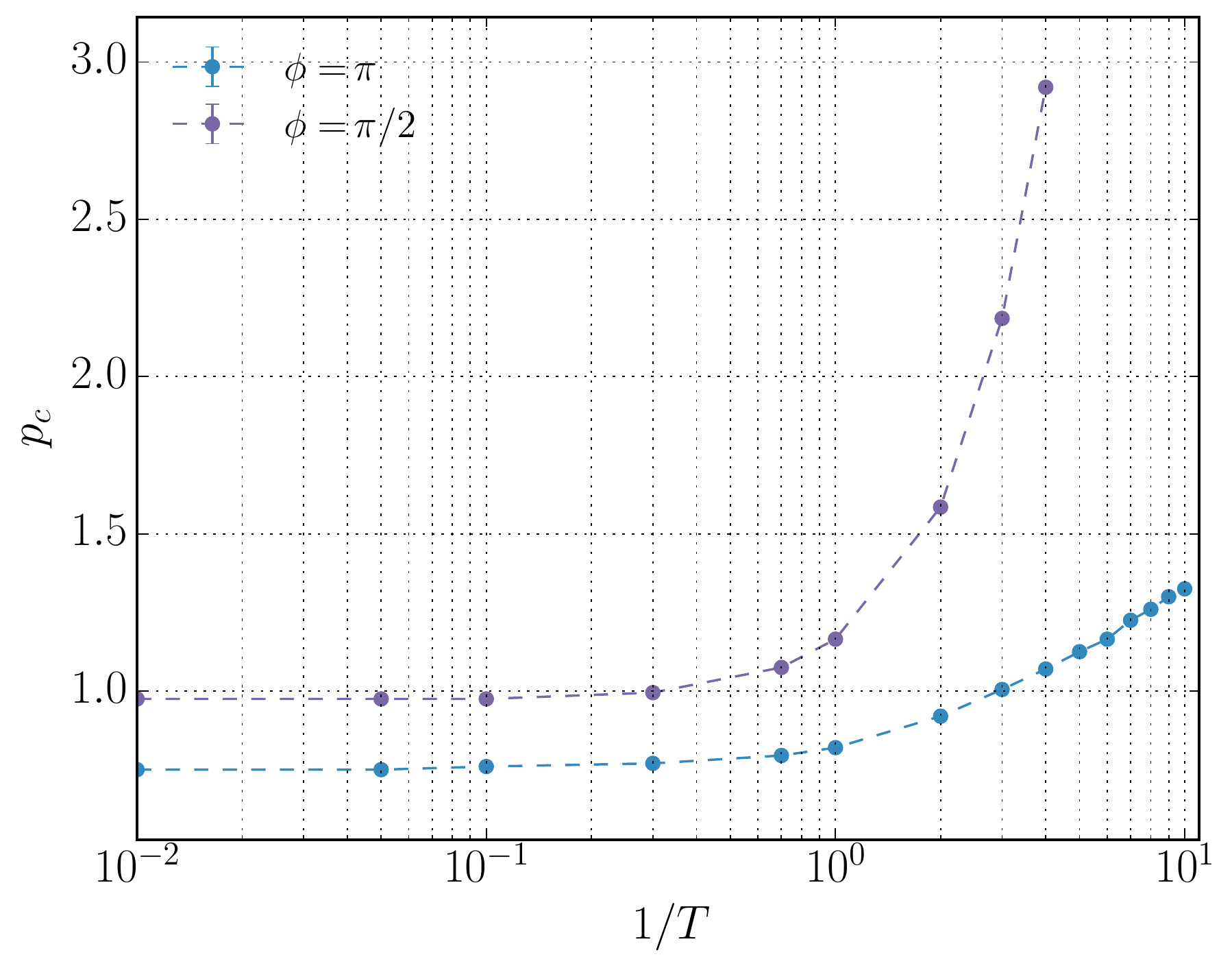}
\includegraphics[width=0.49\columnwidth]{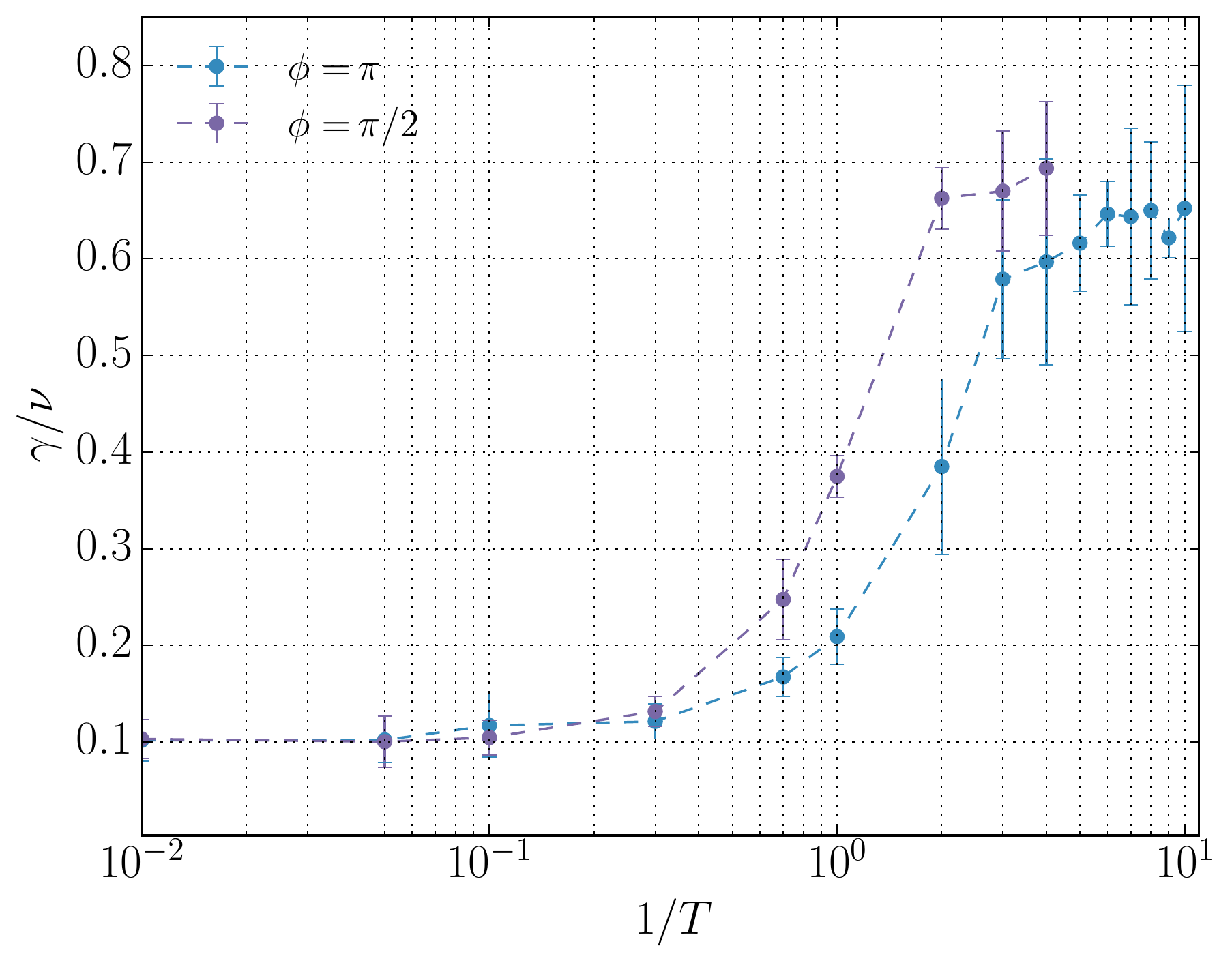}
\caption{\textit{Left:} phase diagram in the plane $(1/T, p_c)$. $p_c$ are estimated with the method of Binder cumulant, averaging every $p$ point over 100 realizations and using $N = 25\, 000, 50\, 000, 100\, 000, 250\, 000, 500\, 000$. The region above the curve corresponds to the phase in which a GSCC appears. The error on $p_c$ is the interval separating $p_c$ from the values of $p$ on its left and right. \textit{Right:} flow of the ratio of critical exponents $\gamma / \nu$. Errors bars span $95\%$ confidence intervals and are computed as errors on linear fits as the one in the right panel of figure \ref{fig:binder_and_gammanu}.}
\label{fig:phase_and_flow}
\end{figure}

The aforementioned analysis can be repeated for a range of values of $1/T$, allowing us to draw a phase diagram using $p$ and $1/T$ as parameters. From the left panel of figure \ref{fig:phase_and_flow} we see that $p_c$ increases as $1/T$ increases. In the region above the curve there is a GSCC component, while in the region below the curve there is no cluster whose size is of order $\mathcal{O}(N)$. The quantity $1/T$ is the inverse temperature measuring the fluctuations around the average direction of spins (which is the x axis). As $1/T$ increases, fluctuations are suppressed and most spin will tend to be aligned. Hence, disks with a larger radius will be needed to form a GSCC. In general, we expect this to be true for the following reason. At $T=0$, it is easy to see that the spins are all oriented in the same direction with probability one. This implies that a disk can be connected only with disks that are ``in front of'' it, within the range of angles $[-\phi/2, +\phi/2]$, along the direction of the external field, meaning that whenever a path from disk $i$ to disk $j$ exists, there will be no path from disk $j$ to disk $i$. Hence, the maximal strongly connected component will have size equal to one. For $T$ small but strictly positive, the probability of having directed paths going back is small but non-zero, implying that one needs a large number of disks, or a large radius, to create paths going back, so that the maximal strongly connected component has a size comparable with the total number of disks.

In the right panel of figure \ref{fig:phase_and_flow} we plot the flow of the ratio $\gamma / \nu$ with respect to $1/T$, for two values of $\phi$. Interestingly, we see that the ratio $\gamma / \nu$ is not constant with respect to $1/T$, meaning that at least one of the critical exponents is changing with the temperature. 
An intuitive motivation of such behavior is the following: $T$ changes the degree of anisotropy of the model by interpolating between a model in which links are formed only along a preferred direction ($T = 0$) and a model in which no such direction exists ($T \to \infty$). For $T \to \infty$ the probability of drawing a given network realization is invariant under spatial rotations. On the contrary, for $T = 0$ this is not the case, as the probability of drawing any configuration with spins sensibly deviating from the preferred direction quickly goes to zero. Hence, by changing $T$ and therefore by gradually breaking a symmetry of the system, we expect the critical behavior of the model to be radically different. Indeed this is what we observe in the right panel of figure\ \ref{fig:phase_and_flow}, in which the ratio $\gamma/\nu$ has different behaviors for small and large values of $T$.

Finally, we can see from figure\ \ref{fig:phase_and_flow} that a smaller value of $\phi$ implies that $p_c$ becomes larger, as it is more more difficult to establish edges in the network. Interestingly, both for small and for values of $T$, the behavior of $\gamma/\nu$ seems not to depend on $\phi$, while in the intermediate region a statistically significant discrepancy can be observed.

Estimating the values of the critical exponents is possible, at least in principle. In fact, once the ratios are known, the knowledge of one exponent, i.e.\ $\nu$ is sufficient to compute all of them. $\nu$ is usually also estimated by means of finite-size scaling. Using the scaling relation \cite{binder} $\chi N^{-\gamma/2\nu} = g(N^{1/2\nu} t)$, where $g$ is a scaling function, we see that at the peak $\chi^*$, $N^{1/2\nu} t$ must be equal to some constant $a$, meaning that: $p^* = p_c + a N^{-1/2\nu}$, where we remind that $p^*$ is the value of $p$ corresponding to $\chi^*$. However, in our simulations the error on $p^*$ turns out to be quite large, being equal to the resolution with which values of $p$ are scanned in the numerical simulations. With the present resolution, the relative error on $\nu$ is easily around $100\%$. Increasing the resolution leads to noisy estimates of $p^*$, which are likely to improve only using a larger number of realizations over which each data point is averaged. In summary, more extensive simulations would be needed for a precise estimation of the critical exponents. Strictly speaking, this difficulty prevents a direct verification of the hyperscaling relation. However, the Binder cumulant evaluated at the critical point does not depend on the system size if and only if the hyperscaling relation holds, which is consistent with the crossing we observe in the left panel of figure \ref{fig:binder_and_gammanu}.


\section{Conclusions} \label{sec:conclu}
In this paper we studied the emergence of a giant strongly connected component in a model of overlapping disks endowed with spin representing the direction along which connections between the centers of disks are formed. The model is inspired by the connectivity properties of neuronal networks, in which axons prolongate to connect to other neurons' dendrites. The motivation for studying the strongly connected component, rather than the connected one, comes from the interest in the subset of neurons that mutually influence each others. In the terminology of machine learning, the strongly connected giant component represents a fully recurrent neural network. 

Our results, albeit restricted to the two-dimensional case, point towards a few interesting facts. First, we find that for a large range of temperatures there exist a critical percolation threshold which we were able to obtain numerically, providing a phase diagram of the model. We note that, while in the case of the undirected networks the density of connections is sufficient to determine whether one is in the percolating phase or not, in the case of directed network one needs to measure not only the density of connections, but also the average orientation of the axons. In addition, we provide strong evidence that the critical exponents change with the temperature, i.e.\ with the parameter controlling the fluctuations around a privileged direction for spins. As a consequence, rather than a single universality class, as we vary the temperature we span a family of universality classes. This feature has been already observed in other models \cite{Noh2004,contperc1,contperc2,contperc3,powerlaw}. 

We believe our simplified model could be of interest in the context of the pruning process taking place in the mature brain, which can be interpreted as a reverse percolation process in which connections are indeed removed. In fact, by sufficiently decreasing the density it is possible to cross the critical line and transition to the phase in which there is not a giant strongly connected component. 

Further investigations are required to understand whether the observed behavior occurs also in more realistic settings, as a three-dimensional spatial embedding and the presence of long-range connections between neurons. The first case, also due to the scaling of the number of disks implied by the Penrose theorem \cite{Penrose2003}, is much more computationally intensive. In the second case, one could follow the approach of small-world networks in which connections are randomly re-wired with a certain (small) probability. However, one should carefully account for the role played by directionality, which could be disrupted by such re-wirings. Intuitively, one would expect that the presence of long-range connections could facilitate the emergence of a GSCC. Finally, as already explained in section \ref{sec:analysis}, a precise estimations of all critical exponents also requires considerable computational resources. We plan to characterize the behavior of all such extensions in future studies.

\ack
F.\ Caravelli would like to thank Leo Kadanoff\footnote{We salute Leo Kadanoff (1937-2015), who passed away during the preparation of this paper, for his important contributions to Physics and for inspiring many generations of statistical physicists.} for various useful comments during this work at the Alumni Conference at Perimeter Institute, and the London Institute of Mathematical Sciences for hospitality. We also would like to thank Rory Finnegan and Luca Dall'Asta for useful comments on the presentation of the results. The work of F.\ Caravelli was supported by Invenia Labs. M.\ Bardoscia acknowledges support from: FET IP Project MULTIPLEX (317532). F.\ Caccioli acknowledges support of the Economic and Social Research Council (ESRC) in  funding  the  Systemic  Risk  Centre (ES/K002309/1).

\end{document}